# Development of a Connected and Automated Vehicle Longitudinal Control Model

**Mizanur Rahman*, Ph.D.**
Postdoctoral Research Fellow
Center for Connected Multimodal Mobility ($C^2M^2$)
Glenn Department of Civil Engineering, Clemson University
125 Lowry Hall, Clemson, SC 29634
Tel: (864) 650-2926; Email: mdr@clemson.edu

**Md Rafiul Islam**
Ph.D. Student
Department of Mathematics and Statistics, Texas Tech University
1108 Memorial Circle, Lubbock, TX 79409
Tel: (806) 317-4096; Email: rafiul.islam@ttu.edu

**Mashrur Chowdhury, Ph.D., P.E., F.ASCE**
Eugene Douglas Mays Endowed Professor of Transportation
Center for Connected Multimodal Mobility ($C^2M^2$)
Glenn Department of Civil Engineering, Clemson University
216 Lowry Hall, Clemson, SC 29631
Tel: (864) 656-3313; Email: mac@clemson.edu

**Taufiquar Khan, Ph.D.**
Professor
School of Mathematical and Statistical Sciences, Clemson University
Martin Hall O-201, Clemson, SC 29634-0975
Tel: (864) 656-3257; Email: khan@clemson.edu

*Corresponding author

**ABSTRACT**

It is envisioned that, in the future, most vehicles on our roadway will be controlled autonomously and will be connected via vehicle-to-everything (V2X) wireless communication networks. Developing a connected and automated vehicle (CAV) longitudinal controller, which will consider safety, comfort and operational efficiency simultaneously, is a challenge. A CAV longitudinal controller is a complex system where a vehicle senses immediate upstream vehicles using its sensors and receives information about its surroundings via wireless connectivity, and move forward accordingly. In this study, we develop an information-aware driver model (IADM) that utilizes information regarding an immediate upstream vehicle of a subject CAV through CAV sensors and V2X connectivity while considering passenger comfort and operational efficiency along with maintaining safety gap for longitudinal vehicle motion of the autonomous vehicle. Unlike existing driver models for longitudinal control, the IADM intelligently fuses data received from in-vehicle sensors, and immediate upstream vehicles of the subject CAV through wireless connectivity, and IADM parameters do not need to be calibrated for different traffic states, such as congested and non-congested traffic conditions. It only requires defining the subject CAV's maximum acceleration and deceleration limit, and computation time that is needed to update the subject CAV's trajectory from its previous state. Our analyses suggest that the IADM (i) is able to maintain safety using a newly defined safe gap function depending on the speed and reaction time of a CAV; (ii) shows local stability and string stability and (iii) provides riding comfort for a range of autonomous driving aggressiveness depending on the passenger preferences. In addition, we conducted a case study, which proves that IADM improves CAV operational efficiency, compared to the existing state-of-the-art Intelligent Driver Model (IDM), without compromising passenger comfort.

**NOTATIONS**

$s^{comm}(t)$ = Reliable wireless communication coverage distance of a subject connected and automated vehicle (CAV) at time $t$

$s^{sens}(t)$ = Reliable in-vehicle sensor coverage distance of a subject CAV at time $t$

$s^{fveh}(t)$ = Gap between a subject CAV and an immediate upstream vehicle of the CAV at time $t$

$s^{safe}(t)$ = Dynamic safe gap between a subject CAV and an immediate upstream vehicle of the CAV at time $t$ so that a CAV can reduce its speed with a comfortable deceleration

$s^{net}(t)$ = Available gap beyond the dynamic safe gap, $s^{safe}(t)$, at time $t$ between a CAV and an immediate upstream vehicle of the CAV at time $t$.

$s^0$ = Safe standstill gap between a CAV and an immediate upstream vehicle of the CAV when the speed of the vehicle is zero at any time $t$

$s^{fgap}(t)$ = Minimum available gap between $s^{comm}(t)$, $s^{sens}(t)$, and $s^{fveh}(t)$ at time $t$

$v^{fveh}$ = Speed of an immediate upstream vehicle of a subject CAV at time $t$

$v^{CAV}(t)$ = Speed of a subject CAV at time $t$

$v^{freeflow}(t)$ = Free flow speed of a roadway section on which a subject CAV is traveling at time $t$

$v^{fspeed}$ = Speed of the immediate upstream vehicle at time $t$ if a vehicle exists within the reliable sensor or communication coverage distance of the subject CAV, and the speed of the subject CAV is constrained by an immediate upstream vehicle, (Or free-flow speed of a roadway at time $t$ if no vehicle exists within the reliable sensor or communication coverage distance of the subject CAV or the speed of the subject CAV is not constraint by an immediate upstream vehicle)

$a^{CAV}(t)$ = Acceleration of a subject CAV at time $t$

$b^{CAV}(t)$ = Deceleration of a subject CAV at time $t$

$a^{comf}(t)$ = Comfortable acceleration of a subject CAV at time $t$

$b^{comf}(t)$ = Comfortable deceleration of a subject CAV at time $t$

$a^{max}$ = Maximum acceleration of a subject CAV

$b^{max}$ = Maximum deceleration of a subject CAV

$\delta$ = Exponent for the vehicle's acceleration

$T$ = Time headway to follow the immediate upstream vehicle of a subject CAV

$i$ = Number of vehicles $(1,2,3,\ldots,N)$

$j$ = Number of observations $(1,2,3,\ldots,M)$ of each vehicle during a selected observation period

# INTRODUCTION

In the coming years, vehicles will be controlled autonomously and will be connected via vehicle-to-everything (V2X) wireless communication networks (e.g., vehicle-to-vehicle (V2V), vehicle-to-infrastructure (V2I), vehicle-to-pedestrians (V2P)) [1]. Ridesharing companies are already undertaking test runs of autonomous taxi services in US and Singapore [2]-[4]. However, functions of a connected and automated vehicle (CAV) controller is complex where a CAV needs to sense the surrounding environment using its sensors and exchange information with other road users and infrastructure using wireless connectivity. In this connected and automated vehicle scenario, a CAV reacts based on the received data and moves as directed by an autonomous vehicle controller software, while maintaining safety, comfort of the passenger and operational efficiency in a mixed traffic environment, which includes all type of road users (automated and non-automated vehicles, connected and non-connected vehicles, and vulnerable road users, such as pedestrians, bicyclists and road workers).

As shown in Figure 1, sensors, V2X communication and actuators are the three blocks of a CAV hardware system, and perception, planning and control are the three blocks of a CAV software system. The perception block fuses information received from CAV sensors and through V2X communication so that the CAVs aware about the surrounding environment and could locate itself. The planning block makes decisions to determine an optimal path and trajectories to reach a destination while avoiding obstacles and optimizing routes. There are three modules of planning block [5] that include: i) mission planning (to determine paths to reach a destination), (ii) behavioral planning (to identify executable tasks for the motion planning block by predicting movement of other participants, and (iii) path planning (to generate optimal trajectory, which will help meet local goals. Control block refers to determining executable control information from the planned actions produced by the planning block.

A CAV scans the surrounding environment to identify the moving and static entities, and choose a safe and efficient path and trajectory to reach a destination using these blocks as shown in Figure 1. However, multiple CAVs can share their sensor, and surrounding environmental data among themselves, which can be transmitted using V2X communication to the perception module of a subject CAV. This shared information includes both sensor-captured data from those cooperative vehicles, and information related to each vehicle's own movements and surrounding environments. Such cooperative movement can enhance the operation of CAVs via vehicle platooning [6], [7]. Information can be shared between surrounding autonomous vehicles using available wireless communication options, such as dedicated short-range communication (DSRC) and 5G, in real-time with minimum delay and maximum reliability [1].

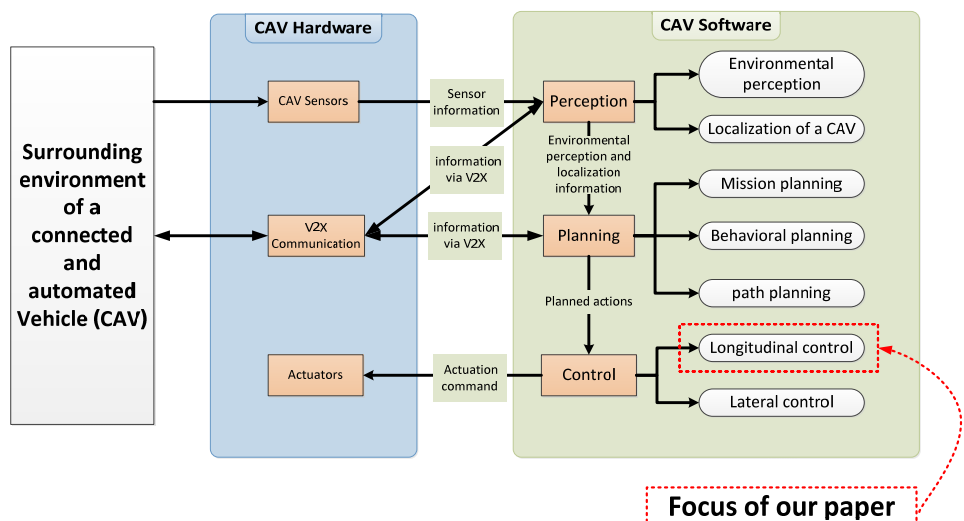

**Figure 1: A high-level connected and automated vehicle (CAV) system architecture and the focus of our paper**

The penetration of automated vehicles will be low in the near future, and in this mixed traffic environment, automated vehicles will move with different road users including human-driven vehicles [1], [8]. The automated vehicle controller needs to actuate its acceleration and deceleration in such a way so that autonomous vehicle movement must be compatible with immediate upstream vehicles, and passengers in the CAV should feel comfortable as it moves through a mixed traffic environment. Driver longitudinal behavior models (also known as car-following models), which can mimic human driving behavior, can be utilized in an automated vehicle controller to have it driven through a mixed traffic environment [1], [9], [10], [11], [12]. As the driver behavior models are necessary for autonomous vehicles for ensuring safe movement, efficient operation and riding comfort of the passengers, the focus of our paper is developing an improved driver model for connected and automated vehicle longitudinal control that can enhance autonomous vehicle safety and operation while maintaining user comfort.

Two types of driver car-following behavior models exist: machine learning-based and kinematics models [13]. Many existing autonomous vehicles have implemented machine learning-based car-following models [14], [15]. These models utilize human-driven car-following data to capture the pattern of human driving maneuvers. Although machine learning based car-following models have the ability to capture human-driven patterns of longitudinal movements with a high accuracy [16]-[22], there are two key disadvantages of these models [18]: (i) the parameters in the models have no physical meanings, and the models need to be calibrated depending on the driving scenarios; and (ii) models need to be trained with massive amounts of data for different possible driving scenarios including unexpected situations. Recent studies concluded that Reinforcement Learning or Supervised Learning could partially solve these issues [18-20]. However, a machine learning based car-following model can make inappropriate decisions if the model does not train with the data for longitudinal movements related to unexpected driving scenarios.

Different from the machine learning based models, kinematics models are not data-driven and these models describe the kinematic mechanisms of car-following maneuvers, such as the GHR model, optimum velocity model and Intelligent Driver Model (IDM) [23]-[27]. Kinematics models have explicit mathematical form and the model parameters have definite physical meanings [23]-[27]. These models can ensure safety and control of longitudinal car-following movements with calibrating model parameters. [18], [23]. However, existing state-of-the-art driver behavior models, such as the IDM and Gipps model, have limitations in terms of operational efficiency and passenger comfort. For example, the IDM model has a slow response for any actuation (acceleration or deceleration), which indicates lower operational efficiency [11]. On the other hand, the Gipps model does not differentiate between maximum acceleration (or deceleration) and comfortable acceleration (or deceleration). The Gipps model only considers safety and it does not consider the comfort of the occupants inside a vehicle [12]. In addition, all models are parameter-dependent and need to be calibrated depending on the driving behavior in different traffic states [23], [12].

Different human drivers have different perception and reaction times, and different drivers will perceive the same object differently, thus driving maneuvers will vary accordingly. In addition, a human driver has a longer reaction time between perceiving and reacting to an object than the controller (software) of an automated vehicle [1]. Computation time threshold for safety [1], [8] in AVs in less than a hundred milliseconds, which includes locating a vehicle using CAV in-vehicle sensors or through data from external sources using V2X communication, and the decision and execution of acceleration and deceleration decision. Furthermore, the riding comfort of an occupant depends on the rate of change of acceleration and deceleration of a vehicle. To confirm passenger comfort, it is required to control the rate of change of acceleration or deceleration, which may reduce the operational efficiency of a CAV. In addition, a vehicle must stop with a comfortable deceleration to ensure safety.

In this study, we develop an information-aware driver model (IADM) that utilizes information through CAV sensors and receive information about immediate upstream vehicles of the subject CAV through V2X connectivity while the CAV controller considers passenger comfort and operational efficiency along with maintaining safety gap for longitudinal vehicle motion. Unlike existing driver models for the longitudinal control, the IADM intelligently fuses data from in-vehicle sensors, and immediate upstream vehicle through wireless connectivity; and model parameters do not need to be calibrated for different traffic states, such as congested and non-congested traffic states. It only requires defining two parameters:

maximum acceleration and deceleration, and computation time that is needed to update the subject CAV's trajectory from its previous state. We analyzed the efficacy of the IADM analytically and numerically to ensure safety and passenger comfort. After that, we conducted a case study to evaluate the IADM in terms of comfort and operational efficiency by simulating a Cooperative Adaptive Cruise Control (CACC) platoon of four vehicles. A CACC scenario with different traffic states has been used to investigate the comfort and operational efficiency of the IADM compared to the existing state-of-the-art Intelligent Driver Model (IDM) model.

**RELATED WORK**

Because of the need for riding comfort of the passengers in a CAV, it is necessary to mimic human driving behavior and incorporate it into a CAV controller design, especially in a mixed traffic scenario through a car-following model. Previously, these models have been used in a traffic micro simulator to mimic human driving. The objective of these models is to replicate the longitudinal driving maneuvers of a subject vehicle's driver while following an immediate upstream vehicle on a roadway [8], [9], [11], [23], [24]. A driver behavior model for following a vehicle has been extensively studied since the 1950s resulting in many models [23]. These include Gazis–Herman–Rothery (GHR) model, linear model, fuzzy-logic based model, collision avoidance model, meta models, optimum velocity model; IDM; Gipps' model and psycho-physical models. For a CAV longitudinal control, a driver model must ensure passenger comfort without compromising operational efficiency in different traffic states [1], [11].

The high number of kinematics car-following models presented their inability to reproduce both traffic flow and vehicle-to-vehicle interactions realistically [23]. As a result, the classical car-following models, such as Gipps' car-following model, has been applied for longitudinal control applications. Gipps model can replicate longitudinal driving behavior for both congested and free-flow traffic states [12], [25]. The maximum acceleration of these traffic conditions is determined based on two constraints: i) the drivers desired speed; and ii) the minimum gap between a subject vehicle and an immediate upstream vehicle that is required to avoid collisions. It (i.e., ii) ensures safe longitudinal movements of a vehicle in a traffic stream. According to Gipps' model, a vehicle that uses Gipps' model can adjust its speed smoothly to reach the desired free-flow speed or safely follow the immediate upstream vehicle. Although the Gipps' model can follow the immediate upstream vehicle closely, it does not differentiate between maximum acceleration (or deceleration) and comfortable acceleration (or deceleration), and the model is unable to show string stability for stop and go traffic scenario [12].

On the other hand, the IDM is another state-of-the-art car-following model, which captures the dynamics of congested and free-flow conditions realistically [11], [12], [26], [27]. In this model, the acceleration of a subject vehicle is the function of the subject vehicle's speed, the ratio of an available gap and the desired gap between subject and the immediate upstream vehicle, and the relative speed between a subject vehicle and an immediate upstream vehicle. The IDM was used to control the longitudinal movements of CACC vehicles for a CACC system design. For example, Milanes and Shladover developed three different control systems to evaluate the performance of CACC controller with the IDM: i) ACC system with field data, ii) CACC system with field data, and iii) CACC systems that uses the IDM [11]. Field experiments were performed with production vehicles to evaluate these three controllers. It was found that CACC vehicles that use the IDM (i.e., iii) provide more comfortable car-following behavior than the other two controllers (i.e., i and ii). However, the IDM model shows a slower response and a large variable distance gap between CACC vehicles. In addition, one needs to calibrate IDM model parameters for different traffic states [28]. Furthermore, existing driver models do not consider how data from different sensors and through wireless V2X connectivity will be used to actuate a vehicle's longitudinal control motion as it is needed for a CAV. Thus, it is necessary to develop a driver model for a CAV's longitudinal control that considers passenger comfort and operational efficiency simultaneously using information from in-vehicles sensors and data from immediate upstream vehicles of the subject CAV.

# INFORMATION AWARE DRIVER MODEL (IADM) FOR LONGITUDINAL CONTROL

## IADM Concept

We assume a CAV longitudinally moves through a mixed traffic environment. A mixed traffic environment includes all types of road users, such as non-automated and automated vehicles, and non-connected and connected vehicles. A CAV senses immediate upstream vehicles either using in-vehicle sensors or utilizing wireless connectivity, or both for longitudinal movements. In order to establish wireless connectivity to exchange information between a CAV and immediate upstream vehicles, a CAV can use a low latency wireless communication option, such as emerging 5G technology or Dedicated Short-Range Communications (DSRC) [1]. A driver model, which controls the longitudinal movement of a CAV, utilizes information from the immediate upstream vehicles to actuate its speed (i.e., acceleration/deceleration).

When a CAV moves on a single traffic lane, the subject CAV adjusts its speed (i.e., accelerate or decelerate) based on the available gap and relative speed between the subject CAV and the immediate upstream vehicle of the CAV. A CAV can only measure the available gap and relative speed of the CAV, if a vehicle is within its sensor or communication range (in the case where a vehicle can send and receive data from other CAVs). Otherwise, a CAV will adjust its speed based on the free flow speed of a roadway if the CAV speed is not constraint by an immediate upstream vehicle or there is no vehicle within the reliable sensor or communication coverage distance. Free-flow speed can be defined as the average speed of the roadway segment at a low-density traffic in which the speed of a CAV will not be constrained by the immediate upstream vehicle of the CAV [29]. For safe operation of the CAV, the available gap will be the minimum of the distance covered by the CAV sensor and wireless communication option if there is no vehicle within the reliable sensor or communication coverage distance. Wireless communication is reliable if a message is guaranteed to reach the subject CAV as intended [30]. For CAV applications, there is a stringent requirement for real-time communication network reliability as a wireless communication network can be affected due to the external environment, such as building, trees and roadway traffic density [31]. The reliability of communication network can be measured by two metrics: packet delivery ratio (i.e., the probability of successfully send a message (or packet) from a sender to a receiver) and distribution of consecutive packet drop (i.e., the probability distribution consecutive packet drops between a sender and a receiver) [30-31]. As our longitudinal driver model considers information from different sources to ensure safe, comfortable and efficient operation of a subject CAV in a mixed traffic scenario, we named our model "Information Aware Driver Model (IADM)."

To provide a better understanding of the IADM concept, Figure 2 presents an example of a CAV operation in a mixed traffic scenario. In this scenario, a non-connected vehicle travels on the same lane in front of a CAV. The immediate upstream vehicle is within both the sensor or communication coverage distance of the subject CAV. For this scenario, we define different gaps related to the subject CAV, and reliable sensor or communication coverage distance. Here we will formulate how to determine the minimum available gap and speed of the immediate upstream vehicle. This speed information of the immediate upstream vehicle will be used to calculate the relative speed between the subject CAV and an immediate upstream vehicle. The following are the notations to describe all the variables and parameters related to the IADM.

$s^{comm}(t)$ = Reliable wireless communication coverage distance of a subject CAV at time $t$

$s^{sens}(t)$ = Reliable in-vehicle sensor coverage distance of a subject CAV at time $t$

$s^{fveh}(t)$ = Gap between a subject CAV and an immediate upstream vehicle of the CAV at time $t$

$s^{safe}(t)$ = Dynamic safe gap between a CAV and an immediate upstream vehicle of the CAV at time $t$ so that a CAV can reduce its speed with a comfortable deceleration

$s^{net}(t)$ = Available gap beyond the dynamic safe gap, $s^{safe}(t)$, at time $t$ between a CAV and an immediate upstream vehicle of the CAV at time $t$.

$s^0$ = Safe standstill gap between a CAV and an immediate upstream vehicle of the CAV when the speed of

the vehicle is zero at any time $t$

$s^{fgap}(t)$ = Minimum available gap between $s^{comm}(t)$, $s^{sens}(t)$, and $s^{fveh}(t)$ at time $t$

$v^{fveh}$ = Speed of an immediate upstream vehicle of a subject CAV at time $t$

$v^{CAV}(t)$ = Speed of a subject CAV at time $t$

$v^{freeflow}(t)$ = Free flow speed of a roadway section on which a subject CAV is traveling at time $t$

$v^{fspeed}$ = Speed of the immediate upstream vehicle at time $t$ if a vehicle exists within the reliable sensor or communication coverage distance of the subject CAV, and the speed of the subject CAV is constrained by an immediate upstream vehicle, (Or free-flow speed of a roadway at time $t$ if no vehicle exists within the reliable sensor or communication coverage distance of the subject CAV or the speed of the subject CAV is not constraint by an immediate upstream vehicle)

$a^{CAV}(t)$ = Acceleration of a subject CAV at time $t$

$b^{CAV}(t)$ = Deceleration of a subject CAV at time $t$

$a^{comf}(t)$ = Comfortable acceleration of a subject CAV at time $t$

$b^{comf}(t)$ = Comfortable deceleration of a subject CAV at time $t$

$a^{max}$ = Maximum acceleration of a subject CAV

$b^{max}$ = Maximum deceleration of a subject CAV

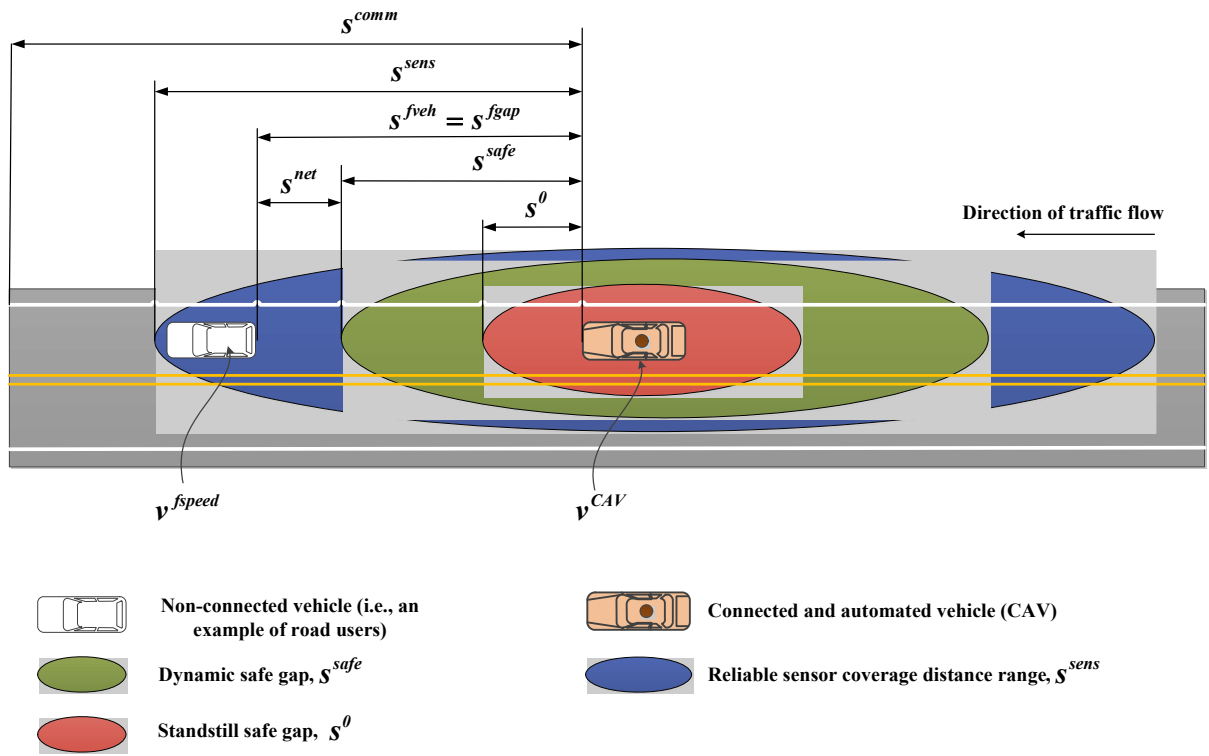

**Figure 2: An example of a connected and automated vehicle (CAV) operation in a mixed traffic scenario within the reliable sensor and communication range, and different types of gaps in front of a CAVs**

As mentioned earlier in this section, a subject CAV uses its sensors (part of autonomous vehicle sensors) or wireless communication (if the subject CAV and its immediate upstream vehicle are wirelessly connected as envisioned in the connected vehicle technology) for speed and location information of the immediate upstream vehicle. Based on the above notations, the equation (1) defines a minimum available gap ($s^{fgap}$) in front of a subject CAV considering reliable wireless communication coverage distance ($s^{comm}$), reliable sensor coverage distance ($s^{sens}$) and gap between a subject CAV and its immediate upstream vehicle ($s^{fveh}$) at time *t*.

$$s^{fgap}(t) = \min\{s^{sens}(t), s^{comm}(t), s^{fveh}(t)\} \tag{1}$$

Similarly, the speed of an immediate upstream vehicle within the reliable sensor or communication coverage distance on the same lane of the CAV can be represented by $v^{fspeed}(t)$ at any time *t*.

$$v^{fspeed} = \begin{cases} v^{fveh} & \text{if the speed of a CAV is constrained by an immediate upstream vehicle} \\ v^{freeflow} & \text{if the speed of a CAV is not constrained by an immediate upstream vehicle} \end{cases}$$

Let's consider that $v^{CAV}(t + \Delta t)$ is the speed of a subject CAV at a time, $(t + \Delta t)$. The acceleration (or deceleration) function of a CAV at time *t* can be defined by equation (2).

$$a^{CAV}(t)\left(or\ b^{CAV}(t)\right) = \lim_{\Delta t \to 0} \frac{v^{CAV}(t+\Delta t) - v^{CAV}(t)}{\Delta t} \tag{2}$$

For a discrete-time model, the time step, $\Delta t$, represents computation time that is needed to update the subject CAV's trajectory from its previous state. IADM considers passenger comfort along with the safety of the CAV vehicles. The passenger comfort will be maintained by keeping the jerk within the comfortable threshold. Jerk can be defined as the rate of change of acceleration (or deceleration). With increasing jerk, passenger comfort will decrease. The IADM incorporates safety, passenger comfort and operational efficiency in a mixed traffic environment by addressing the following considerations:

> **Considerations related to the speed of a CAV:** Let's consider that $v^{CAV}(t + \Delta t)$ is the speed of a subject CAV at a time $(t + \Delta t)$; $a^{comf}(t)$ is the comfortable acceleration of the vehicle at time *t*; $v^{freeflow}(t)$ is the free-flow speed of the roadway at time *t*; $s^{fgap}$ is the available gap between the subject CAV and the immediate upstream vehicle at time *t* and $\Delta t$ is the computation time for updating trajectory of a CAV. The acceleration of a CAV is the function of $s^{fgap}, v^{CAV}$ and $v^{fspeed}$.
>
> The acceleration of a CAV is a strictly decreasing function of its speed, i.e., $\frac{\partial a^{CAV}(t; s^{fgap}, v^{CAV}, v^{fspeed})}{\partial v^{CAV}} < 0$.
>
> A CAV will accelerate to reach the free-flow speed of a roadway if the CAV speed is not constrained by a vehicle or there is no vehicle within the reliable sensor or communication coverage distance (i.e., $s^{fgap} \to \max(s^{comm}, s^{sens})$) of a subject CAV.

$$\lim_{s^{fgap} \to \max(s^{sens}, s^{comm})} a^{CAV}\left(t; s^{fgap}, v^{CAV}, v^{fspeed}\right) = 0 \tag{3}$$

$$\lim_{s^{fgap} \to \max(s^{comm}, s^{sens})} v^{CAV} = v^{freeflow} \tag{4}$$

**Considerations related to the relative speed:** As shown in equation (5), the acceleration of a CAV is a decreasing function of the relative speed between the subject CAV and its immediate upstream vehicle, $\Delta v^{CAV}$, which can be expressed as $(v^{CAV} - v^{fspeed})$.

$$\frac{\partial a^{CAV}(t; s^{fgap}, v^{CAV}, v^{fspeed})}{\partial \Delta v^{CAV}} \leq 0, \tag{5}$$

The acceleration of a subject CAV is an increasing function of the speed of an immediate upstream vehicle as shown in equation (6).

$$\frac{\partial a^{CAV}(t; s^{fgap}, v^{CAV}, v^{fspeed})}{\partial v^{fspeed}} \geq 0 \tag{6}$$

The $v^{fspeed}$ of a CAV is assumed to equal to $v^{freeflow}$ if the CAV speed is not constrained by an immediate upstream vehicle or there is no vehicle within the reliable sensor or communication coverage distance (i.e., $s^{fgap} \to \max(s^{comm}, s^{sens})$) of a subject CAV.

$$\lim_{s^{fgap} \to \max(s^{comm}, s^{sens})} v^{fspeed} = v^{freeflow} \tag{7}$$

$$\lim_{s^{fgap} \to \max(s^{comm}, s^{sens})} \frac{\partial a^{CAV}(t; s^{fgap}, v^{CAV}, v^{fspeed})}{\partial v^{fspeed}} = 0 \quad or,$$

$$\lim_{s^{fgap} \to \max(s^{comm}, s^{sens})} \frac{\partial a^{CAV}(t; s^{fgap}, v^{CAV}, v^{fspeed})}{\partial \Delta v^{CAV}} = 0 \tag{8}$$

**Consideration related to the immediate front gap of a CAV:** The subject CAV always maintains at least, $s^0$, gap from the immediate upstream vehicle. If $s^0 < s^{fgap} < s^{safe}$ i.e., the distance of a CAV from its immediate vehicle smaller than its safe and comfortable distance then a CAV will decelerate with its maximum deceleration to maintain a safe distance or to make a safe stop.

Increasing the gap between the subject CAV and an immediate upstream vehicle, acceleration will increase as the gap increases. The following inequality (i.e., (9)) will be replaced by equality (i.e., (10)) if the immediate upstream vehicle of the subject vehicle is beyond the range of CAV sensors or communication networks.

$$\frac{\partial a^{CAV}(t; s^{fgap}, v^{CAV}, v^{fspeed})}{\partial s^{fgap}} \geq 0 \tag{9}$$

$$\lim_{s^{fgap} \to \max(s^{comm}, s^{sens})} \frac{\partial a^{CAV}(t; s^{fgap}, v^{CAV}, v^{fspeed})}{\partial s^{fgap}} = 0 \tag{10}$$

The detection of an immediate upstream vehicle, and the decision and execution of acceleration (or deceleration) behavior can be done within a hundred milliseconds to assure a safe action. The IADM can mimic a safe, operationally efficient and comfortable driving behavior using the framework of IADM presented here with the acceptable constraint while moving through the city, rural or freeway roadway traffic.

**IADM Formulation**

When a CAV moves through a mixed traffic scenario, a subject CAV perceives immediate upstream vehicles either using its in-vehicle sensor or vehicle-to-everything (V2X) communication or both. We develop a longitudinal motion control model for a CAV controller using information from the in-vehicle sensors and/or V2X communication. If a subject CAV moves through a lane, the CAV only needs information related to the immediate upstream vehicle of the CAV. If there is no vehicle in front of the subject CAV, the subject CAV will follow the roadway free-flow speed. To develop a longitudinal driver behavior model in a mixed traffic scenario, a model must replicate driving behavior that is comfortable to the passenger as it is moving through a mixed traffic scenario, and at the same time, it must assure operational efficiency and safety. Equation (11) represents IADM speed function, $v^{CAV}(t + \Delta t)$, of a subject CAV at time $(t + \Delta t)$, which reflects passenger safety, comfort and operational efficiency.

$$v^{CAV}(t + \Delta t) = \min\{v^{acc}(t + \Delta t),\ v^{freeflow}(t + \Delta t),\ v^{dec}(t + \Delta t)\} \quad (11)$$

Equation (11) calculates three different values for the speed of a subject CAV at time $t + \Delta t$. Three-speed values will be calculated using three-speed function, which includes (i) speed function using comfortable acceleration; (ii) speed function using free-flow speed of a roadway; and (ii) speed function using comfortable deceleration. The minimum of these three values will be used as a speed of the $v^{CAV}$ at time $(t + \Delta t)$.

The speed function for acceleration, $v^{acc}(t + \Delta t)$, can be defined as shown in equation (12).

$$v^{acc}(t + \Delta t) = v^{CAV}(t) + a^{comf}(t) * \Delta t \quad (12)$$

$v^{acc}(t + \Delta t)$ defines a speed function of a CAV at $(t + \Delta t)$ with comfortable acceleration. Later in this section, we defined a comfortable acceleration function ($a^{comf}(t)$) based on the relative speed, relative gap, and choice of autonomous vehicle driving aggressiveness based on the passenger preference.

$v^{dec}(t + \Delta t)$ defines a speed function that calculates the safe and comfortable speed of a CAV at a time $(t + \Delta t)$ as shown in equation (13) depending on the immediately available gap beyond the dynamic safe gap, which is $s^{net}(t)$. $s^{net}(t)$ is the difference between the dynamic safe gap, $s^{safe}(t)$, at time *t* and the minimum available gap, $s^{fgap}(t)$, between a CAV and an immediate upstream vehicle at time *t*. IADM enable acceleration or deceleration to maintain $s^{net}(t)$ equal to zero. In this way, IADM can ensure safety while keeping a safe gap in front of it.

A speed function using comfortable deceleration, $b^{comf}(t)$, of a subject CAV and available gap, $s^{net}(t)$, beyond the safe gap can be defined as follows.

$$v^{dec}(t + \Delta t) = \sqrt{\left(v^{fspeed}(t)\right)^2 - 2b^{comf}(t) * s^{net}(t)} \quad (13)$$

$$s^{net}(t) = s^{fgap}(t) - s^{safe}(t) \quad (14)$$

We calculate a safe gap, $s^{safe}$, at time *t* considering the current speed and relative speed of the subject CAV. The calculated safe gap ensures the safety gap in front of a CAV. The safe gap between a CAV and the immediate upstream vehicle's speed can be defined as:

$$s^{safe}(t) = s_0 + v^{CAV}(t) * \Delta t + \max\{0, (v^{CAV}(t) - v^{fspeed}(t)) * \Delta t\} \quad (15)$$

Our model defines a comfortable acceleration and deceleration based on the relative speed between a subject CAV and its immediate upstream vehicle's speed, and a distance beyond a comfortable safe gap, $(s^{fgap} - s^{safe})$, within the reliable sensor or communication coverage distance of a subject CAV. Equations (16) and (17) define comfortable acceleration and deceleration.

$$a^{comf} = \begin{cases} a^{max} \tanh(k * \mathrm{abs}(v^{fspeed} - v^{CAV})) & \text{if } v^{fspeed} \neq v^{CAV} \quad (16a) \\ a^{max} \tanh(k * \mathrm{abs}(s^{fgap} - s^{safe})) & \text{if } v^{fspeed} = v^{CAV} \quad (16b) \end{cases}$$

$$b^{comf} = -\begin{cases} b^{max} \tanh(k * \text{abs}(v^{fspeed} - v^{CAV})) & if \ v^{fspeed} \neq v^{CAV} \\ b^{max} \tanh(k * \text{abs}(s^{fgap} - s^{safe})) & if \ v^{fspeed} = v^{CAV} \end{cases}$$ (17a)
(17b)

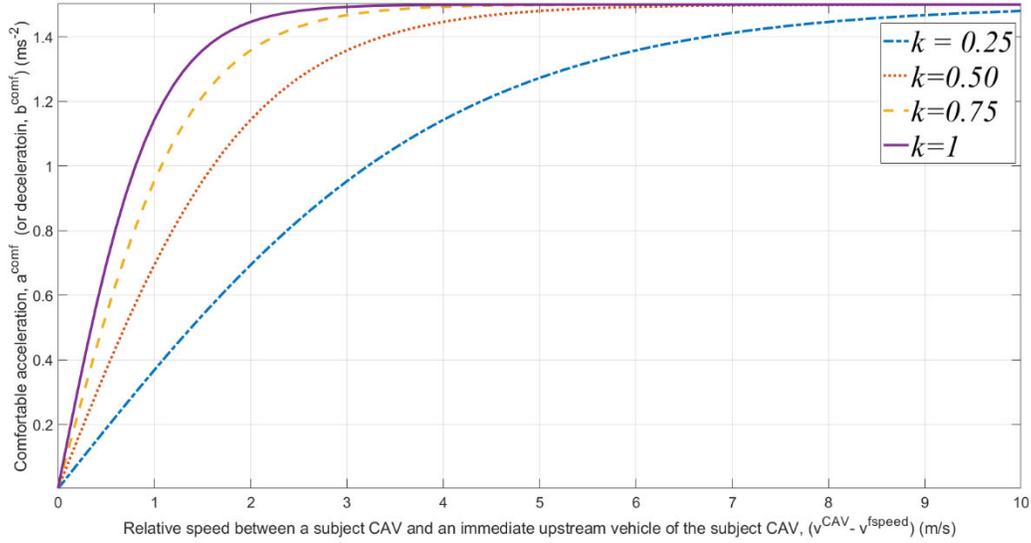

3(a)

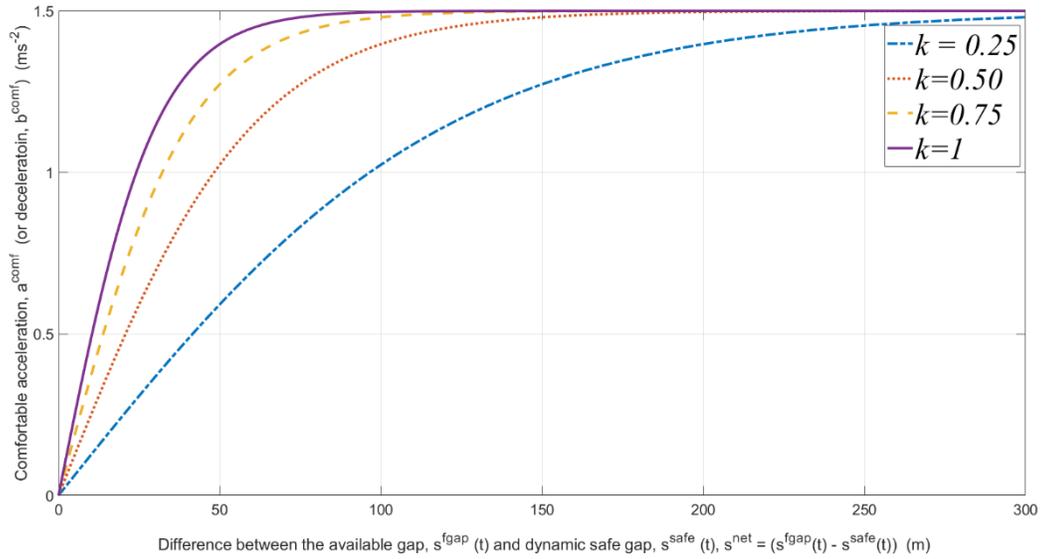

3(b)

**Figure 3: Comfortable acceleration or deceleration by changing (a) relative speed, or (b) difference between the available gap, $s^{fgap}$, and dynamic safe gap, $s^{safe}$, of the subject CAV at any time *t***

Figures 3(a) and 3(b) shows how acceleration and deceleration changes depending on the relative speed of a CAV with respect to its immediate upstream vehicle, $(v^{fspeed} - v^{CAV})$; available gap beyond the dynamic safe gap, $(s^{fgap} - s^{safe})$; and passenger preference related to autonomous driving aggressiveness, which is reflected by passenger type and represented by a factor, *k*. One can control the driving aggressiveness of a CAV using the factor, *k*. Although IADM will assure passenger comfort based on the rate of change of acceleration or deceleration (i.e., jerk) for any value of *k* between 0 and 1, autonomous driving aggressiveness can vary based on the passenger's preferences. In a CAV, a passenger can select their driving aggressiveness preferences through a human-machine interface. If the relative speed is higher than 10, maximum acceleration (or deceleration) will be applied for any value of *k*. However, if *k=1*, maximum acceleration (or deceleration) will be applied after any $s^{net} = (s^{fgap} - s^{safe}) = 120$ m and relative speed 4 m/s. If the rate of change acceleration (or deceleration) increases, the jerk of a CAV will increase and the passenger comfort will decrease. As shown in Figures 3(a) and 3(b), *k=1* curve will give the highest jerk for any change of relative speed between the subject CAV and an immediate upstream vehicle and the available gap beyond the dynamic safe gap, $s^{net} = (s^{fgap} - s^{safe})$; Thus, if the IADM provides acceptable jerk at *k=1*, our model will ensure safe and comfortable operation for any other *k* values.

## IADM Model Equilibria Analysis

The equilibria analyses are necessary to prove the efficacy of the model for the static and dynamic equilibria states. We prove the usefulness of the model theoretically by ensuring safe standstill gap, $s_0$, between a subject CAV and its immediate upstream vehicle of the subject CAV at the static equilibrium state, and the dynamic safe gap, $s^{safe}(t)$, between a subject CAV and an immediate upstream vehicle at time *t* so that a CAV can reduce its speed with a comfortable deceleration at the dynamic equilibrium state. In the following subsections, we present proof of the static and dynamic equilibria of the model.

### *Static Equilibrium*

At the static equilibrium state, the safe gap between a subject CAV and an immediate upstream vehicle at time *t* must be equal to the safe standstill gap at time *t* when the speed of the subject CAV is zero. To prove the static equilibrium condition, we set, $v^{CAV}(t + \Delta t) = 0$, based on the static equilibrium condition. Now we write the equation (11) as follows.

$$v^{CAV}(t + \Delta t) = 0 = \min \{v^{CAV}(t) + a^{comf}(t) * \Delta t, v^{freeflow}(t + \Delta t), \sqrt{(v^{fspeed}(t))^2 - 2b^{comf}(t) * s^{net}(t)}\} \quad (18)$$

As the free-flow speed of a roadway, $v^{freeflow}(t + \Delta t)$, must be positive, we will have the following cases from equation (18).

Case 1: $v^{CAV}(t) + a^{comf}(t) * \Delta t = 0$ (19)

If $v^{CAV}(t) + a^{comf}(t) * \Delta t = 0$, the speed of the subject CAV is given by equation (19) as follows.

$$v^{CAV}(t) = -a^{comf}(t) * \Delta t \quad (20)$$

Equation (20) holds only if $a^{comf}(t)=0$ and $v^{CAV}(t) = 0$ because $v^{CAV}(t) \geq 0$ and $a^{comf}(t) \geq 0$ and $\Delta t > 0$. As shown in equation (16b), $a^{comf}(t) = 0$ holds if the speed of a subject CAV, $v^{CAV}(t)$ equals to speed of its immediate front vehicle, $v^{fspeed}(t)$, and $s^{fgap}(t) = s^{safe}(t)$.

Thus, $v^{fspeed}(t) = v^{CAV}(t) = 0$; and $s^{fgap}(t) = s^{safe}(t) = s^0$. It proves that a CAV maintains the safe standstill gap, $s^0$, between a subject CAV and an immediate upstream vehicle of the subject CAV at any time *t* at the static equilibrium state.

Rahman, Islam, Chowdhury and Khan*Case 2:* $\sqrt{\left(v^{fspeed}(t)\right)^2 - 2b^{comf}(t) * s^{net}(t)} = 0$ (21)

If $\sqrt{\left(v^{fspeed}(t)\right)^2 - 2b^{comf}(t) * s^{net}(t)} = 0$, equation (21) can be written as follows.

$(v^{fspeed}(t))^2 = 2b^{comf}(t) * s^{net}(t)$ (22)

As the comfortable acceleration $b^{comf}(t) \leq 0$, equation (22) holds only if $v^{fspeed}(t) = 0$ and $s^{net}(t) = 0$. As shown in equation (17b), $b^{comf}(t) = 0$ holds if the speed of a subject CAV, $v^{CAV}(t)$ equals to speed of its immediate upstream vehicle, $v^{fspeed}(t)$, and $s^{fgap}(t) = s^{safe}(t)$.

As shown in equation (17), $b^{comf}(t) = 0$ holds only if the speed of a subject CAV, $v^{CAV}(t) = 0$, and speed of its immediate upstream vehicle, $v^{fspeed}(t) = 0$.

Thus, $v^{fspeed}(t) = v^{CAV}(t) = 0$ and $s^{net}(t) = 0$.

From equations (14) and (15), we can write as follows.

$s^{fgap}(t) = s^{safe}(t) = s^0$

Thus, a CAV maintains the safe standstill gap, $s^0$, between a subject CAV and an immediate upstream vehicle of the subject CAV at any time *t* at the static equilibrium state.

From case 1 and case 2, we conclude that the minimum gap between a subject CAV and its immediate upstream vehicle at time *t*, $s^{fgap}(t)$, will be equal to the dynamic safe gap at time *t*, $s^{safe}(t)$ if $v^{fspeed}(t) = v^{CAV}(t) = 0$. So, the safe gap in front of a subject CAV is equal to a safe standstill gap, $s^0$, in front of a CAV at static equilibrium condition.

*Dynamic Equilibrium*
At the dynamic equilibrium state, the safe gap between a subject CAV and an immediate upstream vehicle of the subject CAV at any time *t* must be equal to the dynamic safe gap, $s^{safe}(t)$ so that a CAV can reduce its speed with a comfortable deceleration. For the dynamic equilibrium, $v^{CAV}(t + \Delta t) = v^{CAV}(t)$. Based on this equilibrium condition, we can write equation (11) as follows.

$v^{CAV}(t + \Delta t) = v^{CAV}(t) = min\ \{v^{CAV}(t) + a^{comf}(t) * \Delta t, v^{freeflow}(t + \Delta t),$
$\sqrt{\left(v^{fspeed}(t)\right)^2 - 2b^{comf}(t) * s^{net}(t)}\ \}$ (23)

We will have the following cases from equation (23).

*Case 1:* We set equation (24) based on the dynamic equilibrium condition of equation (23).

$v^{CAV}(t + \Delta t) = v^{CAV}(t) = v^{CAV}(t) + a^{comf}(t) * \Delta t$ (24)

If $v^{CAV}(t + \Delta t) = v^{CAV}(t) = v^{CAV}(t) + a^{comf}(t) * \Delta t$, which yields

$a^{comf}(t) * \Delta t = 0$ (25)

As $\Delta t > 0$, equation (25) holds only if the comfortable acceleration of a CAV, $a^{comf}(t) = 0$. From equation (15), we can write, $v^{fveh}(t) = v^{CAV}(t)$ and $s^{net}(t) = 0$ and equation (15) can be written as follows.

$$s^{fgap}(t) = s^{safe}(t) = s_0 + v^{CAV}(t) * \Delta t$$

*Case 2:* We set equation (26) based on the dynamic equilibrium condition of equation (23).

$$v^{CAV}(t + \Delta t) = v^{CAV}(t) = v^{freeflow}(t + \Delta t) \tag{26}$$

If equation (26) holds, $v^{fspeed}(t) = v^{fspeed}(t + \Delta t) = v^{freeflow}(t + \Delta t)$.

Therefore, the comfortable acceleration of a CAV, $a^{comf}(t) = 0$, and from the equation, we can write as follows.

$$s^{fgap}(t) = s^{safe}(t) = s^0 + v^{CAV}(t) * \Delta t$$

For this case, at the dynamic equilibrium, a subject CAV will either follow its front vehicle's speed, i.e., $v^{CAV}(t) = v^{fspeed}(t)$ or reach the free-flow and it will maintain the following safe distance, $s^{safe}(t) = s^0 + v^{CAV}(t) * \Delta t$.

*Case 3:* We set equation (27) based on the dynamic equilibrium condition of equation (23).

$$v^{CAV}(t + \Delta t) = v^{CAV}(t) = \sqrt{\left(v^{fspeed}(t)\right)^2 - 2b^{comf}(t) * s^{net}(t)} \tag{27}$$

From equation (27), we can write as follows,

$$(v^{CAV}(t))^2 = (v^{fspeed}(t))^2 - 2b^{comf}(t) * s^{net}(t) \tag{28}$$

As per the definition of $b^{comf}(t)$ and equation (28), a subject CAV follows the immediate upstream vehicle's speed, i.e., $v^{fspeed}(t) = v^{CAV}(t)$ if and only if $b^{comf}(t) = 0$ or $s^{net}(t) = 0$.

Based on the equations (14) and (15), the minimum gap in front of a subject CAV at time *t*, $s^{fgap}(t)$, will be equal to the safe gap at time *t*, $s^{safe}(t)$ or reach the free-flow speed and it will maintain the following safe distance.

$$s^{fgap}(t) = s^{safe}(t) = s^0 + v^{CAV}(t) * \Delta t$$

From case 1, case 2 and case 3, we conclude that the available gap, $s^{fgap}(t)$, between a subject CAV and an upstream vehicle of the subject CAV at time *t* is equal to the dynamic safe gap, $s^{safe}(t)$ if $v^{fspeed}(t) = v^{CAV}(t)$ at the dynamic equilibrium state.

**SIMULATION AND NUMERICAL ANALYSIS OF IADM**

We evaluated the efficacy of the IADM in terms of safety and passenger comfort by simulating a CACC platoon of four vehicles. In a CACC platoon, each vehicle is connected and automated. We assumed no data loss when each vehicle of the CACC platoon wirelessly communicate with each other, and we considered 100ms communication delay, which is the standard vehicle-to-vehicle communication latency for any safety applications [21].

**Simulation and Numerical Analysis Scenario**

We simulated three different traffic states to assess driver car-following behaviors of vehicles in a CACC platoon. Three traffic states include (i) uniform speed state, (ii) speed with constant acceleration state, and (iii) speed with constant deceleration state. Table 1 presents numerical analysis scenarios and parameter values of IADM. A total number of vehicles for forming a CACC platoon, total simulation time, initial position and speed of the follower and leader vehicles, position and speed profile of a vehicle immediately upstream of the leader vehicle of the CACC platoon and IADM parameter values are the input parameters for the analysis. In the simulation, the number of vehicles of the platoon including follower vehicles and platoon leader was four. The initial position of an immediate upstream vehicle of the platoon leader was 80 m from the origin and its initial speed was 15 m/s. The initial position of the platoon of four vehicles including leader was 60 m, 40 m, 20 m and 0 m from the origin. The initial speed of all follower vehicles was 15 m/s, which was same as the leader vehicle's speed. The speed of all the CACC vehicles varies between 15m/s (~35mph) to 25m/s (~ 55 mph), which represents the traffic flow characteristics of the urban roadway. The authors divide the total simulation durations into seven periods: (i) uniform speed state (15 m/s), (ii) constant acceleration state (speed changes from 15 m/s to 25 m/s), (iii) uniform speed state (25 m/s), (iv) constant deceleration state (speed changes from 25 m/s to 15 m/s), (v) uniform speed state (15 m/s), (vi) constant acceleration state (speed changes from 15 m/s to 20 m/s), and vii) uniform speed state (20 m/s). For the evaluation of the performance of the IADM model for a CACC controller design in this study, we use $a^{max}=1.5\ ms^{-2}$, $b^{max}=-1.5\ ms^{-2}$; length of each vehicle=5.0m and free-flow speed of a roadway, $v^{freeflow}=25$ m/s.

**Table 1 Simulation and numerical analysis scenarios and parameter values of IADM**

| Input Parameters | Parameter values |
|---|---|
| Total vehicle number | One leader and three follower vehicles |
| Initial position of the immediate upstream vehicle of the CACC platoon leader | 80 m from the origin |
| Initial position of the leader Vehicle of the CACC platoon | 60 m from the origin |
| Initial position of follower Vehicles (1, 2 and 3) | 40 m, 20 m and 0 m respectively from the origin |
| Initial speed of the immediate upstream vehicle of the CACC platoon leader | 15 m/s |
| Initial speed of the vehicles in the CACC platoon (1, 2, 3 and 4) | 15 m/s for all follower vehicles |
| Maximum acceleration | $a^{max}=1.5\ ms^{-2}$ |
| Maximum deceleration | $b^{max}=-1.5\ ms^{-2}$ |
| Simulation time | 200 s |
| Driver aggressiveness factor, $k$ | 1 (for IADM) |
| Traffic states defined by the vehicle immediately upstream of the leader vehicle of the CACC platoon | • **Uniform speed state** (15 m/s, 20 m/s and 25 m/s)<br>• **Speed with constant acceleration state** (Speed changes from 15 m/s to 25 m/s and from 15 m/s to 20 m/s)<br>• **Speed with constant deceleration state** (Speed changes from 25 m/s to 15 m/s) |

**Analysis of Safety**

Using the numerical analysis described in the previous section, Figure 4 was created to show the speed profiles of a vehicle in front of a platoon leader of a CACC platoon and each vehicle of a CACC platoon of the four CAVs that uses IADM for their longitudinal control. IADM shows that each CAV closely follows the speed of the platoon leader CAV with the IADM CACC controller model. Figure 5 presents gap profiles for each vehicle of the CACC platoon that uses IADM. A positive gap indicates that there is no collision between vehicles that use IADM while maintaining a higher speed. Gaps for all CACC follower vehicles are similar to each other for a certain speed, which indicates that there is insignificant fluctuations between the immediate gap of follower vehicles within the CACC in different traffic states. Consistency between gaps of all follower vehicles that use IADM implies that there is no shockwave propagating among CAVs when the platoon is moving from one traffic state to another traffic state. However, the gap cannot represent passenger comfort.

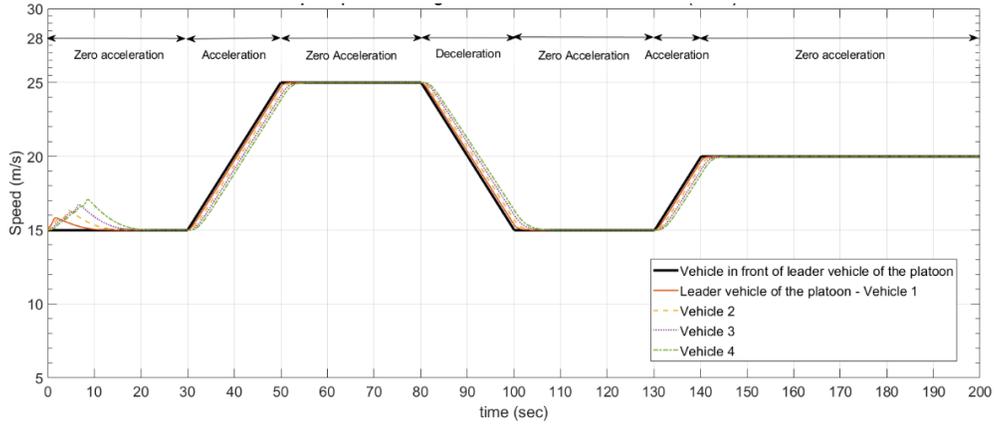

**Figure 4: Speed profile of each vehicle in a CACC platoon of four CAVs that use IADM**

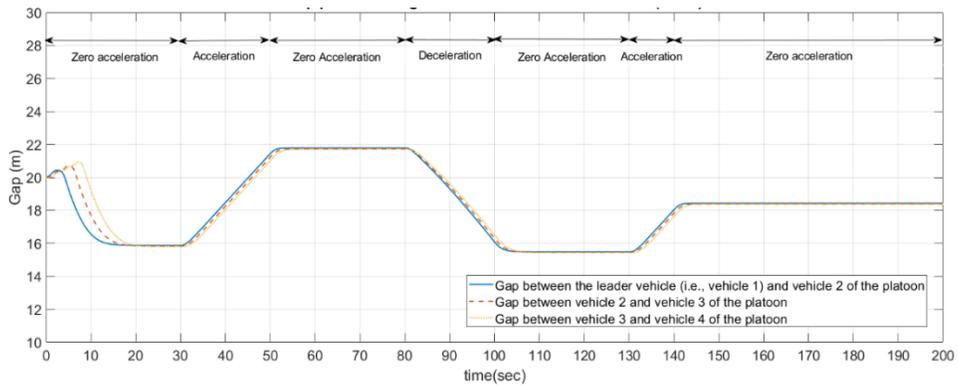

**Figure 5: Gap between a subject CAV and an immediate upstream vehicle of a CACC platoon of four vehicles that use IADM**

**Analysis of Passenger Comfort**

Figure 6 presents the acceleration (or deceleration) profile of each CAV in the CACC platoon of four vehicles that uses IADM for longitudinal control. Although IADM is showing smooth acceleration (or deceleration) profiles in Figure 6, it is not possible to evaluate how comfortable the ride is by only observing the acceleration (or deceleration) profiles. It is also necessary to investigate jerk to analyze the level of passenger comfort. Jerk, which represents the rate of change of acceleration (or deceleration), also indicative of passenger comfort. With an increase in jerk, passenger comfort will decrease [32-33].

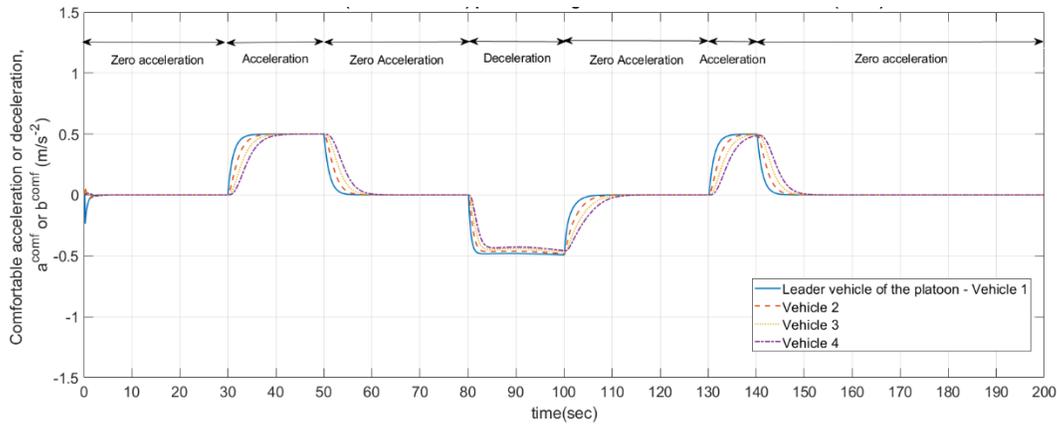

**Figure 6: Acceleration/deceleration profile of each CAV in a CACC platoon of four vehicles that use IADM**

Wei et al. [32-33] used ±2 m/s³ as an acceptable jerk range while ±1 m/s³ as a comfortable jerk range for a passenger, which we will use as a reference in this study as acceptable threshold for passenger comfort. Figure 7 shows that IADM only shows jerk if there is any change of traffic state, such as a change from uniform speed to constant acceleration state or vice versa, and from uniform speed to constant deceleration state or vice versa, otherwise there is no jerk while a CAV is in constant speed, constant acceleration or constant deceleration states. However, the first vehicle in the CACC platoon is showing the maximum jerk, which is less than 1 ms⁻³, and jerk for other vehicles in the CACC platoon is lower than the first vehicle of the platoon instead of being higher. Thus, it indicates that the ride of a passenger will be comfortable (as shown in Figure 7).

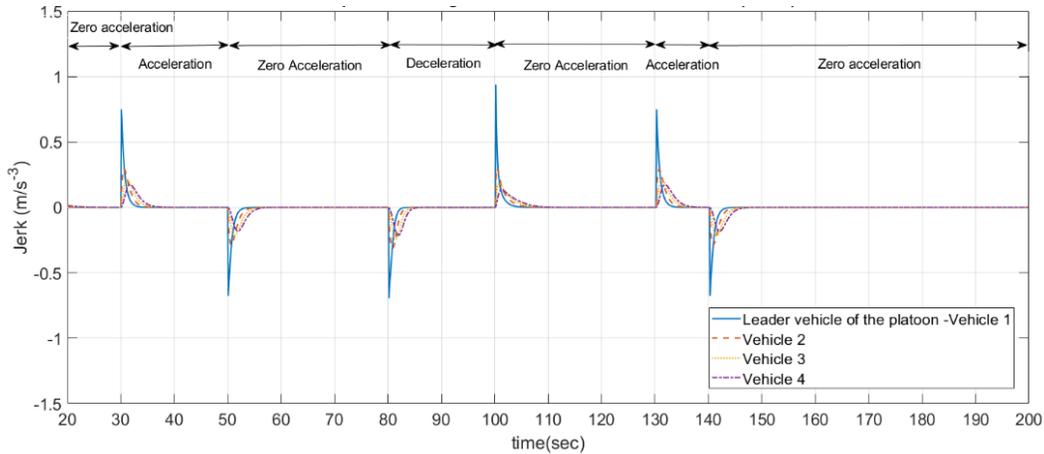

**Figure 7: Jerk profile of each CAV in a CACC platoon of four vehicles that use IADM**

**Local Stability and String Stability Analysis**

A driver behavior model must confirm local stability and string stability. Local instability can be investigated numerically by introducing a perturbation in the speed profile of the immediate upstream vehicle of the CACC platoon leader to have a sudden drop or rise of the speed when a vehicle is following a leader vehicle. A system will be locally unstable if the gap between a subject CAV and an immediate upstream vehicle, and speed fluctuations of the follower vehicles in a CACC platoon increase or do not decrease over time. On the other hand, string stability refers to the stability of a platoon of vehicles. A platoon is string stable if local perturbation in the speed of the leader vehicle in a CACC platoon decreases for all follower vehicles. It is a collective stability phenomenon of finite platoon size. To investigate the local and sting stability, we introduce a perturbation in the speed profile of the immediate upstream vehicle of the CACC platoon leader and then observe the fluctuation of speed profiles of the followers in the CACC platoon. Figure 8 presents the speed profile of each vehicle in a CACC platoon of four CAVs that use IADM. To quantify the speed and gap fluctuation between CAVs, we plotted speed error and gap profiles of the follower CAVs that use IADM with respect to the leader vehicle of the platoon in Figures 9 and 10, respectively. We observe that IADM shows local stability and string stability as the speed, speed error and gap fluctuations do not only increase, and it decreases over time.

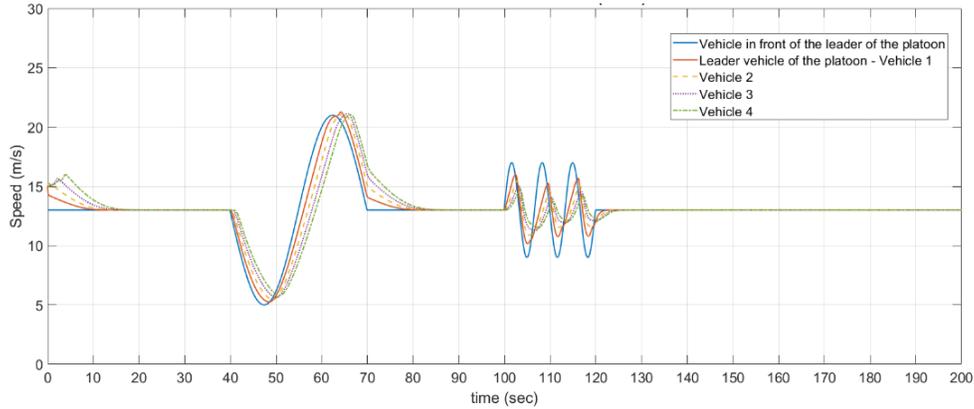

**Figure 8: Speed profile of each CAV in a CACC platoon of four vehicles that use IADM introducing perturbations in the leader speed profile**

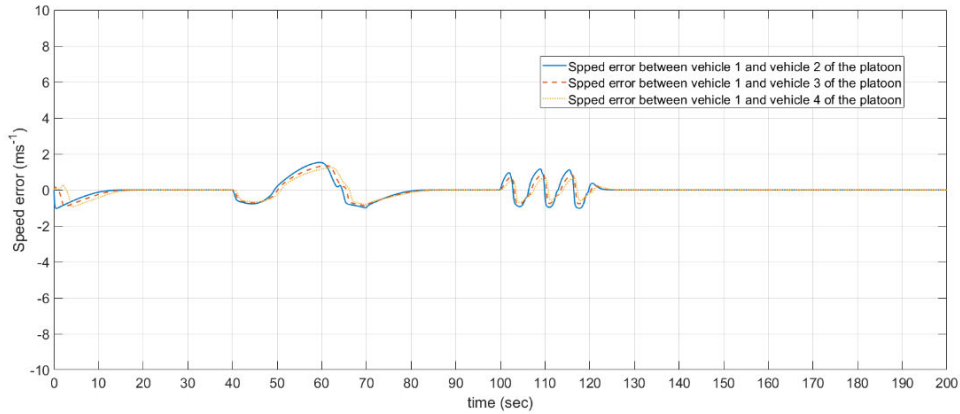

**Figure 9: Speed error profiles of the vehicles of a CACC platoon that use IADM**

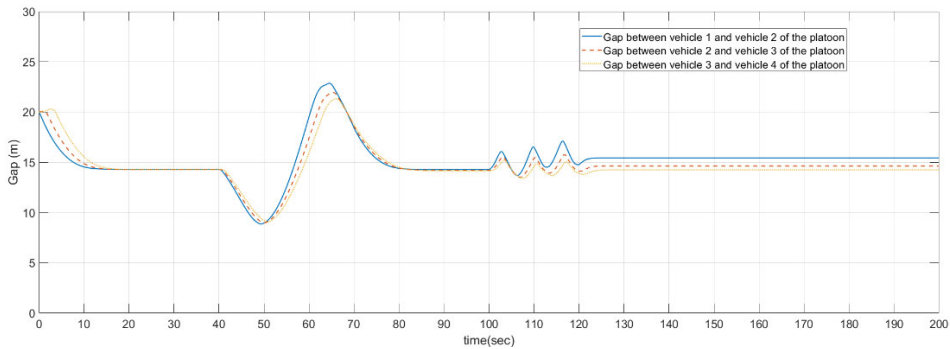

**Figure 10: Gap between a subject vehicle and an immediate upstream vehicle of a CACC platoon**

**CASE STUDY: COMPARISON BETWEEN IADM AND STATE-OF-THE-ART IDM MODEL**

The purpose of this case study is to show an application of the IADM model in a CACC platoon and compare its performance in CACC that uses IDM model. In this case study, we investigate the comfort and operational efficiency of each vehicle in the platoon along with local stability and string stability of IADM and IDM. IDM is the state-art-of-the driver model [8], [12], which can be used as a longitudinal control model for a CAV controller. Thus, we compare the performance of our IADM to IDM in terms of passenger comfort and operational efficiency using the same simulation scenario, which we have used for simulation and numerical analysis of the IADM. As presented in Table 1, we have simulated three different traffic states (i.e., uniform speed state, speed with constant acceleration state, and speed with constant deceleration state) for a simulation period of 200s for this case study.

**IDM Model for CAV Controller and Its Parameter Values**

Equation (29) presents the acceleration (or deceleration) of a subject CAV as per IDM [26, 27]. We can formulate the IDM acceleration equation for the V2X supported CAV controller in the following way.

$$a^{CAV}(t + \Delta t) = a^{max} \left[ 1 - \left( \frac{v^{CAV}(t)}{v^{freeflow}(t)} \right)^\delta - \left( \frac{s^{safe}(t)}{s^{fgap}(t)} \right)^2 \right] \qquad (29)$$

Where,

$v^{CAV}(t) =$ speed of the subject CAV at time t

$a^{max} =$ maximum acceleration of a subject CAV

$v^{freeflow}(t) =$ free-flow speed of a roadway at time *t*

$s^{safe}(t) =$ Safe gap between a subject CAV and an immediate upstream vehicle at time *t*

$s^{fgap}(t) =$ available gap in front of a subject CAV at time *t* (as shown in Figure 2)

$\delta =$ exponent for the vehicle's acceleration

The relative speed of the subject vehicle with respect to the immediate upstream vehicle is as follows:

$$\Delta v^{CAV}(t) = v^{CAV}(t) - v^{fspeed}(t)$$

The dynamic safe gap, $s^{safe}$ at time *t* is defined as follows:

$$s^{safe}(t) = s^0 + max\{0, v^{CAV}(t) * T + \frac{v^{CAV}(t) \Delta v^{CAV}(t)}{2 \sqrt{(ab)}}\} \qquad (30)$$

where,

$b^{max} =$ normal comfortable braking deceleration

$T =$ time headway to follow the immediate upstream of a subject CAV

$$s^{net}(t) = s^{fgap}(t) - s^{safe}(t) \qquad (31)$$

Although IDM is in the form of an ordinary differential equation, IADM is modeled in the form of a difference equation. To compare the performance evaluation of the IADM model with the IDM, we convert the IDM ordinary differential equation into a difference equation form as shown in the equation (32):

$$v^{CAV}(t + \Delta t) = v^{CAV}(t) + \Delta t * a^{max} \left[ 1 - \left( \frac{v^{CAV}(t)}{v^{freeflow}(t)} \right)^\delta - \left( \frac{s^{safe}(t)}{s^{fgap}(t)} \right)^2 \right] \qquad (32)$$

*Rahman, Islam, Chowdhury and Khan*

The authors used the IDM model parameter values estimated in [25] for the evaluation of the performance of the IDM model for a CACC controller design in this study. IDM parameter values are presented in Table 2.

**Table 2 IDM parameters summary** [34]

| Parameter | Value |
|---|---|
| Free flow speed when driving on a road, $v^{splim}$ | 25.0 m/s |
| Standstill safe distance, $s^0$ | 2.0 m |
| Desired safety time headway when following other vehicles, $T$ | 0.1 s |
| Acceleration, $a$ | 1.5 m/s² |
| Braking deceleration, $b^{max}$ | 1.5 m/s² |
| Acceleration exponent, $\delta$ | 4 |

**Analysis of Operational Efficiency**

Figure 11 shows the speed profiles of a vehicle in front of a platoon leader and each vehicle of a CACC platoon of four CAVs that use IDM. As shown in Figure 4 in the "Simulation and Numerical Analysis of IADM" section, IADM shows that each CAV closely follows the speed of the platoon leader CAV with the IADM CACC controller model compared to IDM. Figure 12(a) and 12(b) shows the speed error profiles of each vehicle of the platoon that uses IADM and IDM, respectively, which quantify how closely each follower vehicle follows the leader vehicle's speed. We have calculated the speed error for each follower vehicle (as shown in Figure 12(a)) with respect to the leader vehicle of the platoon, which is vehicle 1.

(speed error of a follower vehicle)$_{i,j}$ = (speed of vehicle 1 at time $i$) − (speed of vehicle $j$ at time $i$)

where,

$i$ represents the number of vehicles $(1,2,3,…,N)$

$j$ = Number of observations $(1,2,3,…,M)$ of each vehicle during a selected observation period

IADM speed error varies between 0 m/s and 2 m/s. On the other hand, IDM speed error varies between 0 m/s and 4 m/s. This clearly indicates that IADM will provide a higher flow rate (i.e., the number of vehicles crossing a certain point per hour on a road) than IDM. We have calculated the sum of $l_1$ and $l_2$ error as provided in Table 3 using the following formulae. Using this speed error, we calculated the $l_1$ norm (equation (33)), which represents least absolute error, and $l_2$ norm (equation (34)), which represents the least square errors, for all follower vehicles. We calculated both $l_1$ and $l_2$ norms as $l_1$ norm can distinguish error between speed profiles even if there is a small deviation between the speeds of the vehicles, and $l_2$ norm can capture the least square error between the speeds of the vehicles.

$$\text{Sum of } l_1 \text{ error} = \sum_{i=1}^{N} \sum_{j=1}^{M} abs\big((\text{speed error})_{i,j}\big) \tag{33}$$

$$\text{Sum of } l_2 \text{ error} = \sum_{i=1}^{N} \left(\sum_{j=1}^{M} \big((\text{speed error})_{i,j}\big)^2\right)^{1/2} \tag{34}$$

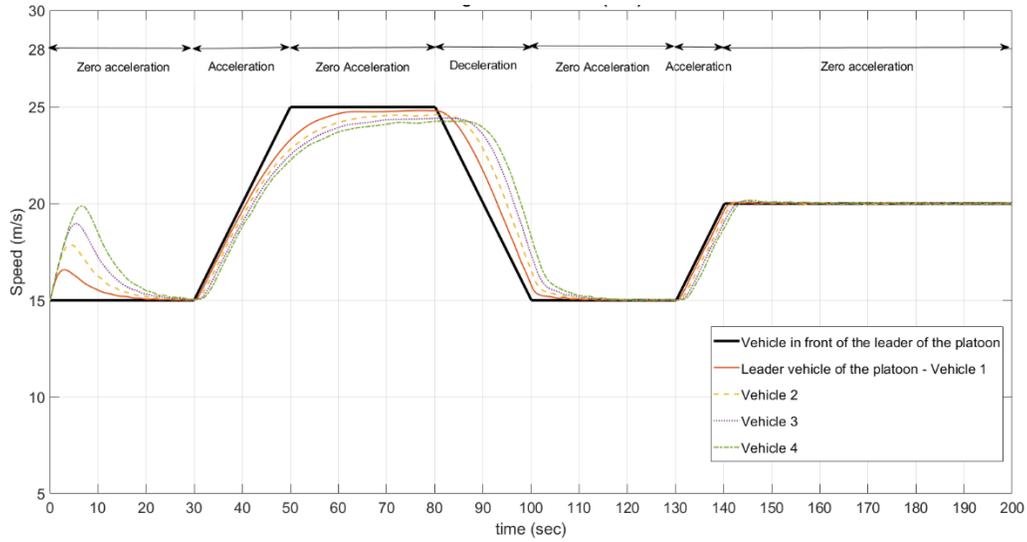

**Figure 11: Speed profile of a CACC platoon of four vehicles that use IDM**

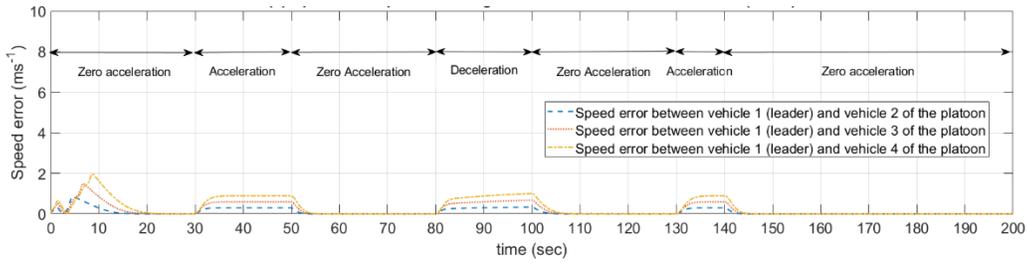

12(a) Speed error profiles using IADM

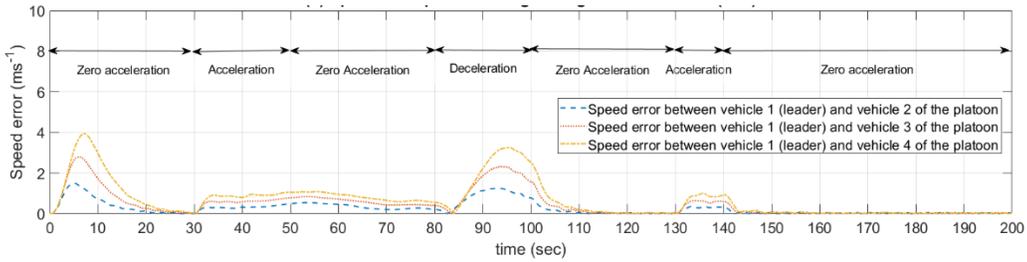

12(b) Speed error profiles using IDM

**Figure 12: Speed error profiles of a CACC platoon of four vehicles that use IADM and IDM**

As shown in Table 3, the $l_1$ speed error for IADM is 1205 while corresponding speed error is 2897 for IDM. On the other hand, the $l_2$ speed error for IADM is 47 while the corresponding speed error is 105 for IDM. The $l_1$ norm that uses Manhattan distance represents least absolute error, and $l_2$ norm that uses Euclidian distance represents least square error. As Euclidean distance calculates the shortest or minimum distance between two points compared to Manhattan distance, $l_1$ error is always higher than $l_2$ error. Thus, $l_1$ error can distinguish error between speed profiles even if there is a small deviation between the speeds of the vehicles. As IADM shows less $l_1$ and $l_2$ errors compared to IDM, it numerically indicates that CAVs in the platoon follows the leader vehicle speed closer than the IDM. However, the gap in front of a CAV could change over time depending on the speed of the subject CAV. As gaps of CAVs that use IADM for all CACC follower vehicles are similar at certain times compared to IDM, consistency between gaps of all IADM follower vehicles indicates the stability of the traffic flow when the platoon is moving from one traffic state to another traffic state. We have calculated $l_1$ and $l_2$ gap error in Table 4. The $l_1$ error for IADM is 782 while error for IDM is 4714. On the other hand, the $l_2$ error for IADM is 39 while error value for IDM is 152. As IADM shows less $l_1$ and $l_2$ errors compared to IDM, it numerically indicates that CAVs of the platoon that use IADM follow the leader CAV of the CACC platoon speed closer than the IDM, which represents the better efficiency of the IADM compared to IDM. The following subsection analyzes passenger comfort of the IADM in terms of rate of change of acceleration and deceleration (i.e., jerk).

**Table 3: Summary of speed error**

| Model Name | $l_1$ error | $l_2$ error |
|---|---|---|
| | Speed error | |
| IADM | 1205 | 47 |
| IDM | 2897 | 105 |

**Table 4: Summary of gap error**

| Model Name | $l_1$ error | $l_2$ error |
|---|---|---|
| | Gap error | |
| IADM | 782 | 39 |
| IDM | 4714 | 152 |

**Analysis of Passenger Comfort**
Figure 13 presents an acceleration/deceleration profile with relative gap and relative velocity of the first following vehicle in a CACC platoon that uses IADM and IDM and it shows smoother acceleration/deceleration behavior, which eventually reduces the jerk and increases riding comfort of a follower CAV that uses IADM compared to a CAV that uses IDM. Figure 14 (a) shows that IADM only shows jerk if there is any change of traffic state, such as a change from uniform speed to constant acceleration state or vice versa, and from uniform speed to constant deceleration state or vice versa, otherwise there is no jerk while a CAV is in constant speed, constant acceleration or constant deceleration states. On the other hand, although IDM provides less jerk, it changes throughout simulation time (as shown in Figure 14(b)). As first 20s is the platoon forming state, we do consider the first 20s of the simulation and we evaluated jerk profile from 20 s to 200s.

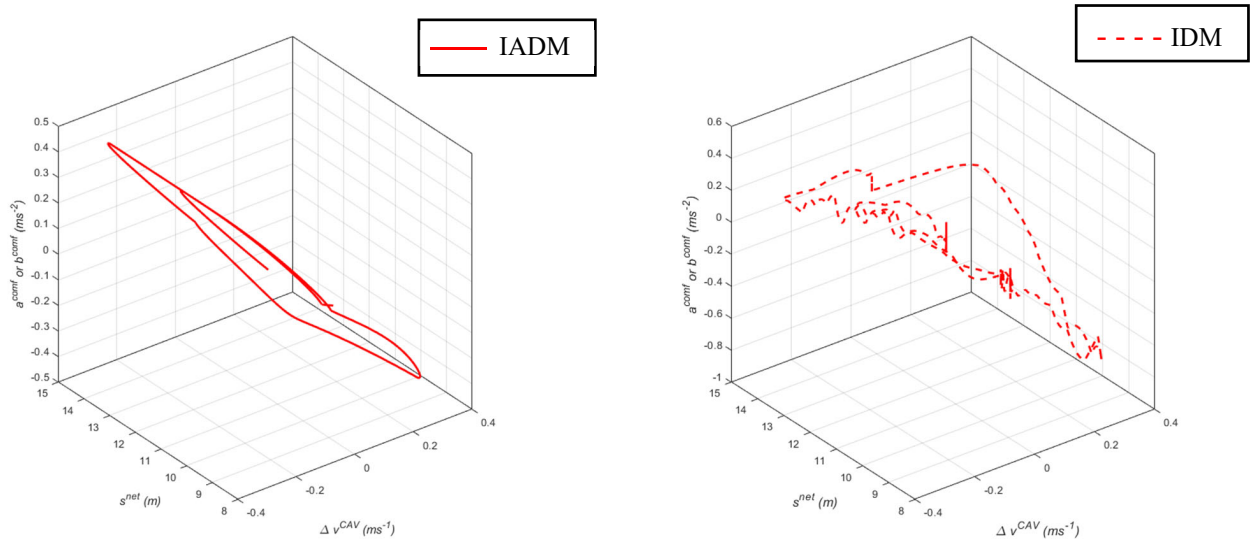

**Figure 13: Acceleration/deceleration profile with relative gap and relative velocity of the first following CAV in a CACC platoon that uses IADM and IDM**

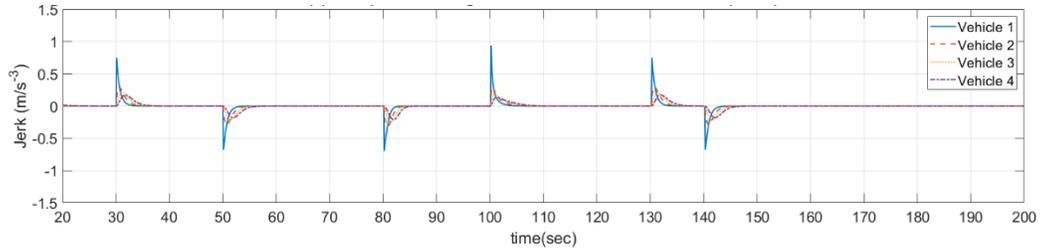

14(a) Jerk profiles using IADM

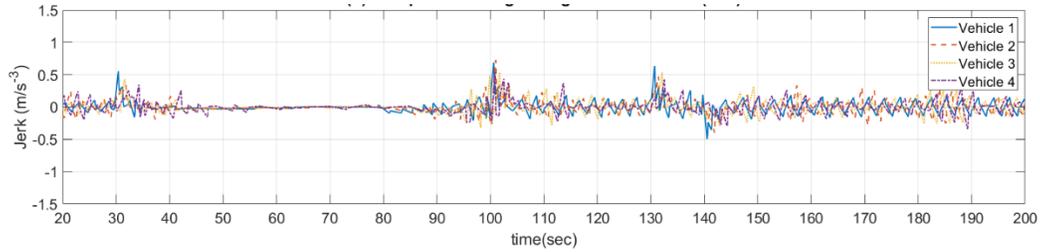

14(b) Jerk profiles using IDM

**Figure 14: Jerk profile of each CAV in a CACC platoon of four vehicles that use IADM and IDM**

## CONCLUSIONS

In this research, we develop an IADM that would consider passenger comfort and operational efficiency for longitudinal vehicle motions of a CAV controller. Our analyses suggest that the IADM is able to maintain safety using a newly defined dynamic safe gap function that depends on the speed and reaction time of a CAV, which provide local stability and string stability as well as provide riding comfort for a range of autonomous driving aggressiveness by formulating different level of acceleration and deceleration functions. To evaluate passengers' comfort, we have analyzed jerk and found that IADM only shows jerk if there is any change of traffic state, such as a change from uniform speed to constant acceleration state or vice versa and from uniform speed to constant deceleration state or vice versa; otherwise, there is no jerk while a CAV is in constant speed, constant acceleration or constant deceleration states. For CAVs that use IADM, jerk is less than 1 ms$^{-3}$, which ensures passenger comfort. In addition, IADM confirms the local stability and string stability as the speed fluctuations do not increase and decay over time. We have also conducted a case study, which shows that IADM improves CAV operational efficiency compared to the Intelligent Driver Model (IDM) without compromising passenger comfort. We found that IADM will provide a higher flow rate than IDM as vehicle speeds in IADM are higher than IDM. The $l_1$ and $l_2$ error quantification indicates a significant improvement in IADM in terms of traffic flow efficiency compared to IDM. We also evaluated gap profiles and found that using IADM, each follower vehicle maintains a similar gap for a certain speed, which is not the case in IDM. Overall, our analyses show that IADM improves operational efficiency while maintaining safety and passenger comfort simultaneously better than IDM. The constrained imposed for the proposed IADM provides the necessary smoothness in the acceleration and speed profiles, which make the IADM a robust model for longitudinal control for CAVs than IDM.

## ACKNOWLEDGMENT

This material is based on a study partially supported by the Center for Connected Multimodal Mobility (C$^2$M$^2$) (USDOT Tier 1 University Transportation Center) Grant headquartered at Clemson University, Clemson, South Carolina, USA. Any opinions, findings, and conclusions or recommendations expressed in this material are those of the author(s) and do not necessarily reflect the views of the Center for Connected Multimodal Mobility (C$^2$M$^2$), and the U.S. Government assumes no liability for the contents or use thereof.

**REFERENCES**

1. Sarker A, Shen H, Rahman M, Chowdhury M, Dey K, Li F, Wang Y, Narman HS. A Review of Sensing and Communication, Human Factors, and Controller Aspects for Information-Aware Connected and Automated Vehicles. IEEE Transactions on Intelligent Transportation Systems, DOI: 10.1109/TITS.2019.2892399, 2019 Mar 15.
2. Zurschmeide, J. Ubers pittsburgh robo taxis amuse riders, still struggle with double parked cars. Available online at: http://www.digitaltrends.com/cars/uber-pittsburgh-robo-taxi-experiment/, Accessed: December 31, 2019.
3. Grenoble, R. The worlds first self-driving taxi fleet is prepping for a test drive. Available online at: http://www.huffingtonpost.com/entry/lyft-self-driving-taxi\us\572b6ab6e4b096e9f0907294. Accessed: December 31, 2019.
4. Ross, P. E. Delphi to test self-driving taxi service in Singapore. Available online at: http://spectrum.ieee.org/cars-that-think/transportation/self-driving/delphi-to-test-robotaxi-service-in-singapore. Accessed: December 31, 2019.
5. González, D., Pérez, J., Milanés, V. and Nashashibi, F., 2015. A review of motion planning techniques for automated vehicles. *IEEE Transactions on Intelligent Transportation Systems*, *17*(4), pp.1135-1145.
6. Cantas, M.R., Gelbal, S.Y., Guvenc, L. and Guvenc, B.A., 2019. Cooperative Adaptive Cruise Control Design and Implementation (No. 2019-01-0496). SAE Technical Paper.
7. Rahman, M., Chowdhury, M., Dey, K., Islam, M.R. and Khan, T., 2017. Evaluation of driver car-following behavior models for cooperative adaptive cruise control systems. Transportation Research Record, 2622(1), pp.84-95.
8. Talebpour A, Mahmassani HS. Influence of connected and autonomous vehicles on traffic flow stability and throughput. Transportation Research Part C: Emerging Technologies. 2016 Oct 1;71:143-63.
9. Treiber, M. and Kesting, A., 2017. The intelligent driver model with stochasticity-new insights into traffic flow oscillations. Transportation research part B, 23, pp.174-187.
10. Sharma, A., Zheng, Z., Bhaskar, A. and Haque, M.M., 2019. Modelling car-following behaviour of connected vehicles with a focus on driver compliance. Transportation Research Part B: Methodological, 126, pp.256-279.
11. Milanés, V., and S. E. Shladover., 2014. Modeling Cooperative and Autonomous Adaptive Cruise Control Dynamic Responses using Experimental Data. Transportation Research Part C: Emerging Technologies, 48, pp 285-300.
12. Treiber, M. and Kesting, A., 2013. Traffic flow dynamics. Traffic Flow Dynamics: Data, Models and Simulation, Springer-Verlag Berlin Heidelberg.
13. Yang, D., Zhu, L., Liu, Y., Wu, D. and Ran, B., 2018. A Novel Car-Following Control Model Combining Machine Learning and Kinematics Models for Automated Vehicles. IEEE Transactions on Intelligent Transportation Systems, *20*(6), pp.1991-2000.
14. Markoff, J. Google cars drive themselves in traffic. New York Times, vol. 10, p. 9, May 2011.
15. Here's How Close Tesla Is to Making True Self-Driving Cars a Reality. Available online at https://observer.com/2019/08/tesla-self-driving-cars-update-andrej-karpathy-ai-director/. Accessed: December 31, 2019.
16. Wang, X., Jiang, R., Li, L., Lin, Y., Zheng, X. and Wang, F.Y., 2017. Capturing car-following behaviors by deep learning. IEEE Transactions on Intelligent Transportation Systems, *19*(3), pp.910-920.

17. Wewerinke, P. Modeling human learning involved in car driving. In the Proc. IEEE Int. Conf. Syst., Man, Cybern., Hum., Inf. Technol., Oct. 1994, pp. 1968–1973.
18. Wei, D., F. Chen, and T. Zhang. 2010. Least square-support vector regression based car-following model with sparse sample selection," in Proc. 8th World Congr. Intell. Control Automat. (WCICA), pp. 1701–1707.
19. Khodayari, A., A. Ghaffari, R. Kazemi, and R. Braunstingl, "A modified car-following model based on a neural network model of the humandriver effects,"IEEE Trans. Syst., Man, Cybern. A, Syst. Humans,vol. 42, no. 6, pp. 1440–1449, Nov. 2012.
20. Colombaroni, C., and G. Fusco. 2014. Artificial neural network models for car following: Experimental analysis and calibration issues. Journal of Intell. Transp. Syst., vol. 18, pp. 5–16.
21. Wei, D., and H. Liu. 2013. Analysis of asymmetric driving behavior using a self-learning approach," Transp. Res. B, Methodol., vol. 47, pp. 1–14.
22. Papathanasopoulou, V. and C. Antoniou. 2015. Towards data-driven car-following models. Transp. Res. C, Emerg. Technol., vol. 55,pp. 496–509.
23. Brackstone, M., and M. McDonald. 1999. Car-Following: A Historical Review. In *Transportation Research Part F: Traffic Psychology and Behaviour*, Vol. 2, No. 4, pp. 181-196.
24. Dey, K.C., L. Yan, X. Wang, Y. Wang, H. Shen, M. Chowdhury, L. Yu, C. Qiu, and V. Soundararaj. 2016. A Review of Communication, Driver Characteristics, and Controls Aspects of Cooperative Adaptive Cruise Control (CACC). IEEE Transactions on Intelligent Transportation Systems, 17(2), pp.491-509.
25. Gipps, P.G., 1981. A behavioural car-following model for computer simulation. Transportation Research Part B: Methodological, 15(2), pp.105-111.
26. Treiber, M., A. Hennecke and D. Helbing. 2000. Congested Traffic States in Empirical Observations and Microscopic Simulations. *Physical Review E* 62, no. 2, pp 1805.
27. Treiber, M., A. Hennecke, and D. Helbing. 2000. Microscopic Simulation of Congested Traffic. *Traffic and granular flow'99*, pp. 365-376. *Springer Berlin Heidelberg*, 2000.
28. Rahman, M., Chowdhury, M., Khan, T. and Bhavsar, P., 2015. Improving the efficacy of car-following models with a new stochastic parameter estimation and calibration method. IEEE Transactions on Intelligent Transportation Systems, 16(5), pp.2687-2699.
29. Manual, Highway Capacity, 2010. Transportation Research Board of the National Academies, Washington, DC, 2010.
30. Bai, F. and Krishnan, H. 2006. September. Reliability analysis of DSRC wireless communication for vehicle safety applications. In 2006 IEEE intelligent transportation systems conference (pp. 355-362). IEEE.
31. Deif, D. and Gadallah, Y., 2017. A comprehensive wireless sensor network reliability metric for critical Internet of Things applications. EURASIP Journal on Wireless Communications and Networking, 2017(1), p.145.
32. Wei, X. and Rizzoni, G., 2004. *Objective metrics of fuel economy, performance and driveability-A review* (No. 2004-01-1338). SAE Technical Paper.
33. Nandi, A.K., Chakraborty, D. and Vaz, W., 2015. Design of a comfortable optimal driving strategy for electric vehicles using multi-objective optimization. *Journal of Power Sources*, *283*, pp.1-18.
34. Gunawan, F. E. Two-vehicle Dynamics of the Car-Following Models on Realistic Driving Condition. *Journal of Transportation Systems Engineering and Information Technology* 12, no. 2, 2012, pp 67-75.